# Constraining Coupling Constants' Variation with Supernovae, Quasars, and GRBs

Rajendra P. Gupta*
*Department of Physics, University of Ottawa, Ottawa, Canada K1N 6N5*

**ABSTRACT**

Dirac, in 1937 proposed the potential variation of coupling constants derived from his large number hypothesis. Efforts have continued since then to constrain their variation by various methods, including astrophysical and cosmological observations. We briefly discuss several methods used for the purpose while focusing primarily on the use of supernovae type 1a, quasars, and gamma-ray bursts as cosmological probes for determining cosmological distances. Supernovae type Ia (SNeIa) are considered the best standard candles since their intrinsic luminosity can be determined precisely from their light curves. However, they have only been observed up to about redshift $z = 2.3$, mostly at $z \leq 1.5$. Quasars are the brightest non-transient cosmic sources in the Universe. They have been observed up to $z = 7.5$. Certain types of quasars can be calibrated well enough for their use as standard candles but with a higher degree of uncertainty in their intrinsic luminosity than the SNeIa. Gamma-ray bursts (GRBs) are even brighter than quasars, and they have been observed up to $z = 9.4$. They are sources of highly transient radiation lasting from 10s of milliseconds to several minutes and, in rare cases, for a few hours. However, they are even more challenging to calibrate as standard candles than quasars. Both the quasars and the GRBs use SNeIa for distance calibration. What if the standard candles' intrinsic luminosities are affected when the coupling constants become dynamic and depend on measured distances? Assuming it to be constant at all cosmic distances would lead to the wrong constraint on the data-fitted model parameters. This paper uses our earlier finding that the speed of light $c$, the gravitational constant $G$, the Planck constant $h$, and the Boltzmann constant $k$ vary such that their variation is interrelated as $G \sim c^3 \sim h^3 \sim k^{3/2}$ with $\dot{G}/G = 3\dot{c}/c = 3\dot{h}/h = 1.5\dot{k}/k = 3.90(\pm 0.04) \times 10^{-10} \text{ yr}^{-1}$ and corroborates it with SNeIa, quasars, and GRBs observational data. Additionally, we show that this covarying coupling constant model may be better than the standard ΛCDM model for using quasars and GRBs as standard candles and predict that the mass of the GRBs scales with $z$ as $((1 + z)^{1/3} - 1)$. Noether's symmetry on the coupling constants is now transferred effectively to the constant in the function relating to their variation.

## 1. INTRODUCTION

Constancy of fundamental constants has been investigated since time immemorial but attained prominence when Dirac in 1937 [1] developed his large number hypothesis, and derived from it that the gravitational constant $G$ and the fine structure constant $\alpha$ may evolve with cosmological time. Teller [2] considered the Solar luminosity dependence on $G$ based on the stellar scaling laws and constrained its possible variation by ensuring that the Solar luminosity was always conducive for life to evolve on Earth over the time of its existence. Most methods developed since then determined the possible variation of $G$ well below Dirac's prediction The methods include the study of solar luminosity evolution [e.g., 3], occultation and eclipses of the Moon [4], evidence based on paleontological data [5], cooling and pulsation of white dwarfs [e.g., 6], evolution of the star clusters [7], masses and ages of neutron stars [8], anisotropies observed in cosmic microwave background radiation [e.g., 9], abundance of light elements from big-bang nucleosynthesis [e.g., 10], asteroseismology data analysis [11], lunar laser ranging observations [e.g., 12], planetary orbits evolution over time [e.g., 13], binary pulsars observations [e.g., 14], evolution of supernovae type-1a (SNeIa) luminosity [e.g., 15], and the observation of gravitational-wave from binary neutron stars [16].

*email: rgupta4@uottawa.ca

Despite Einstein developing the radical theory of special relativity assuming the speed of light $c$ to be *constant*, he also consider it to be *variational* [17]. Qi et al. [18] reported a negligible $c$ variation (assuming a power law variation), and using the observational data for very low and moderate redshift $z$ values of baryon acoustic oscillations (BAO), supernovae type Ia, cosmic microwave background, and Hubble parameter $H(z)$. Another possibility for measuring the $c$ variation with cosmic time was suggested by Salzano et al. [19]. They determined constraints on the variation of $c$ using the relation between the maximum value of the angular diameter distance $D_A(z)$ and $H(z)$, and the BAO and simulated data. Using the independent determination by Suzuki et al. [21] of $H(z)$ and the luminosity distance $D_L(z)$ from SNeIa observations, Cai et al. [20] tried to study the variation of $c$. The first measurement of $c$ value with respect to $z = 1.7$ was reported by Cao et al. [22] and found it essentially the same at $z = 0$, i.e., as measured on Earth. They used the measurement available for radio quasars for angular diameter distance extending to high redshifts. Using galactic-scale strong gravitational lensing systems with quasars / SNeIa as lensed sources, Cao et al. [23] considered a direct measurement of the variation of the speed of light. Lee [24] did a statistical analysis of a galaxy-scale strong gravitational lensing sample that included stellar velocity dispersion measurement on 161 systems and determined effectively no variation in the speed of light. Mendonca et al. [25] obtained negative results in their attempt of determining the variation of $c$ using mass fraction measurements in galaxy cluster gas.

The Planck constant $h$ and the Boltzmann constant $k$ are other constants of interest in our work. The effect of time-dependent stochastic fluctuations of the Planck constant was studied by Mangano et al. [26]. de Gosson [27] applied the effect of the Planck constant's variation on mixed quantum states. The possibility of temporal and spatial variation of the Planck constant was considered Dannenberg [28] by raising it to the status dynamical field that couples to itself and other fields, the coupling being through the Lagrangian density derivatives. He further studied the cosmological implications of such variations and reviewed the literature on the subject. Doppler broadening of absorption lines in thermal equilibrium [e.g., 29, 30] can be used for direct measurement of the Boltzmann constant. The broadening affects the profile of rovibrational line (e.g., of ammonia) along a laser beam. The profile is determined by the Maxwell-Boltzmann molecular velocity distribution which is related to the kinetic energy of each molecule. A critical analysis of spectral line profiles of distant objects, such as quasars and interstellar media, should therefore, in principle, be able to constrain the variation of the Boltzmann constant.

Uzan [31, 32] has comprehensively reviewed the variation of fundamental constants. The validity of the variability of dimensioned vs. dimensionless constants has been of concern to Uzan [31, 32] and others [e.g., 33].

All the experiments and observations known to us have attempted to determine the variation of one constant with all others held invariant. Such an approach may be considered flawed when several constants may be varying in the expressions used for studying their data, especially when variation of one constant may be related to another constant. We have attempted to permit concurrent variation of $c, G, h$, and $k$ in our studies - cosmological, astrophysical, and astrometric: $G \sim c^3 \sim h^3 \sim k^{3/2}$. As a result, using our *covarying coupling constants* (**CCC**) approach[1], we were able to:

*Resolve the primordial lithium problem*: The most widely accepted Lambda Cold Dark Matter (ΛCDM) model yields the Big Bang abundance of lithium about four times higher than observed. Our CCC model calculates it within the uncertainties of the observations [34].

*Find a reasonable solution to the faint young Sun problem*: Stellar evolution models determine the luminosity of the Sun was 25% lower 4.6 billion years ago than today. Thus life could not have evolved as water would have been frozen. The CCC model yields the luminosity to be lower only by 6% [35].

*Prove that gravitational lensing cannot determine the variation of $c$*: A superova Ia light curve has a well defined peak. Since path travelled are different for different gravitationally lesnsed images of the source, the time difference in the peaking of the luminosity between two gravitationally lensed images of a SNIa should in principle determine if $c$ was different at the time light rays were bent by the lens compared to its value today. However, both the geometrical time delay and the Shapiro time delay scale as $G/c^3$, this method is not capable of determining the variation of $c$ [36].

*Establish that supernovae 1a SNeIa data are consistent with the CCC model*: We have shown that the luminosity of the best standard candles SNeIa used for determining large galactic distances is not constant over cosmic time scale. By applying the luminosity correction, the CCC model fits the Pantheon SNeIa data as well as the ΛCDM model [37, 38].

*Verify that quasars can be used to extend Hubble diagram to high redshifts, $z > 7$*: The 'extreme Population A (xA) quasars' approach, sometimes exceed the Eddington luminosity limit. These quasars have the potential to serve as standard candles to measure distances at which supernovae type Ia are too dim to be observable. By establishing how their luminosities vary when coupling constants vary over cosmic time and calibrating them using SNeIa standard candles, we show that xA quasars can be used reliably for measuring such distances [39],

*Determine phenomenologically the covarying relation $G \sim c^3 \sim h^3 \sim k^{3/2}$*: We have determined the scaling law by applying local energy conservation laws to an exploding star – core collapsed supernova – due to runaway nuclear fusion. The fusion energy release converts to the kinetic energy of the explosion after countering its gravitational binding energy. The former partially converts to thermal energy and radiation. I show that when the coupling constants are allowed to vary, they must vary as $G \sim c^3 \sim h^3 \sim k^{3/2}$ [40].

---

[1] We unorthodoxically call the $c, G, h$, and $k$ coupling constants as they determine the strength of different energies involved in a system (mass energy, gravitational energy, photon energy, thermal energy, etc.) which are all coupled to one another. In the context of this paper and for the economy of words these are the only constants we call coupling constants in the CCC model.

*Show from the first principle that $G \sim c^3$*: We have considered a scalar-tensor theory of gravity with the dynamical scalar field $\phi$ comprising $G$ as well as $c$. Both $G$ and $c$ are allowed to be functions of the spacetime coordinates against only $G$ in Brans-Dicke theory. When the system gets to the stable point, the dynamics of $\phi$ cease. Then the constraint $\dot{G}/G = \sigma(\dot{c}/c)$ with $\sigma = 3$ has to be satisfied for the rest of the cosmic evolution [41].

*Demonstrate that constraining one coupling constant leads to constraining the others*: The CCC framework is a modified gravity setup assuming Einstein Field Equations wherein the quantities $\{G, c, \Lambda\}$ are promoted to spacetime functions. We used the ansatz $\dot{G}/G = \sigma(\dot{c}/c)$ with $\sigma =$ constant to deduce the functional forms of $c = c(z)$ and $G = G(z)$. We then showed that this varying $\{G, c, \Lambda\}$ model fits SNeIa data and $H(z)$ data with $\sigma = 3$. [42].

*Show that orbital timing observations do not constrain the variation of $G$*: When $c$ is permitted to vary, and distance is measured using the speed of light, we have shown using the CCC approach that the measured constraints are on the variation of $G/c^3$ and not on $G$ [43].

*Conclude that Kibble balance could be used to measure the variation of the constants*: A Kibble balance measures the gravitational mass of a test mass with extreme precision by balancing the gravitational pull on the test mass against the electromagnetic lift force, resulting in equations leading to mass measurements involving the coupling constants. [44].

The predicted variation of the coupling constants is shown in Figure 1.

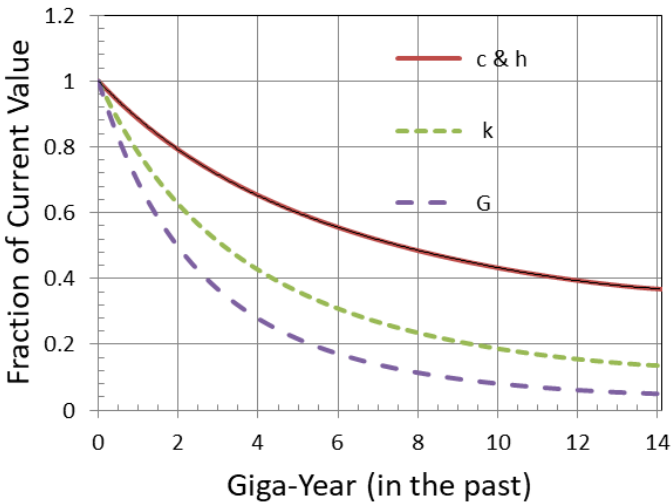

**FIGURE 1**
Variation of $c, G, h,$ and $k$ with cosmic time.

We examine the basic physics of covarying coupling constants (CCC) in Section 2. Section 3 is devoted to the application of CCC to Pantheon supernovae type 1a data [45]. In Section 4, we take the quasars' Hubble diagram data from Marziani & Sulentic [46] and mass evolution data from Vestergaard & Osler [47] and see how our model fits the same cosmological parameters that fit the Pantheon data. Section 5 examines the Hubble diagram data of gamma-ray bursts from Escamilla-Rivera et al. [48, 49], again using the cosmological parameters determined by fitting the supernovae type 1a data. We discuss our findings in Sections 6, and summarize our conclusion in Section 7.

## 2. PHYSICS OF COVARYING COUPLING CONSTANTS

Following Costa et al. [50] we could write the Einstein equations for homogeneous and isotropic universe with coupling constants varying with respect to time $t$ – the speed of light $c = c(t)$, the gravitational constant $G = G(t)$ and the cosmological constant $\Lambda = \Lambda(t)$, as follows:

$$G_{\mu\nu} = \left(\frac{8\pi G(t)}{c(t)^4}\right) T_{\mu\nu} - \Lambda(t) g_{\mu\nu}. \tag{1}$$

Here $G_{\mu\nu} = R_{\mu\nu} - \frac{1}{2} g_{\mu\nu} R$ is the Einstein tensor - $R_{\mu\nu}$ is the Ricci tensor and $R$ the Ricci scalar, and $T_{\mu\nu}$ is the energy-momentum tensor. When we apply the contracted Bianchi identities, local conservation laws, and torsion free continuity,

$$\nabla^\mu G_{\mu\nu} = 0 \text{ and } \nabla^\mu T^{\mu\nu} = 0, \tag{2}$$

We get a general constraint equation for the variation of the coupling constants

$$\left[\frac{1}{G}\partial_\mu G - \frac{4}{c}\partial_\mu c\right]\left(\frac{8\pi G}{c^4}\right) T^{\mu\nu} - (\partial_\mu \Lambda) g^{\mu\nu} = 0. \tag{3}$$

Now we may write the Friedmann–Lemaître–Robertson–Walker (FLRW) metric for the geometry of the universe as:

$$ds^2 = -c^2(t)dt^2 + a^2(t)\left[\frac{dr^2}{1-kr^2} + r^2(d\theta^2 + \sin^2\theta d\phi^2)\right]. \tag{4}$$

Here $k$ (written in units of the inverse of the curvature of the Universe $R^2$) depends on the spatial geometry of the universe: $k = -1$ for negatively curved universe, $k = 0$ for flat universe, and $k = +1$ for positively curved universe.

The energy-momentum tensor (also called stress-energy tensor), may be written as:

$$T^{\mu\nu} = \frac{1}{c^2(t)}(\varepsilon + p)U^\mu U^\nu + p g^{\mu\nu}, \tag{5}$$

assuming that the universe contents can be treated as perfect fluid. Here $\varepsilon$ is the energy density, $p$ is the pressure, $U^\mu$ is the 4-velocity vector having constraint $g_{\mu\nu}U^\mu U^\nu = -c^2(t)$. (We will often drop showing $t$ variation, e.g. $c(t)$ is written as $c$.)

Solving the Einstein equation yields CCC compliant Friedmann equations:

$$H^2 \equiv \frac{\dot{a}^2}{a^2} = \frac{8\pi G\varepsilon}{3c^2} + \frac{\Lambda c^2}{3} - \frac{kc^2}{a^2}, \Rightarrow \dot{a}^2 = a^2\left(\frac{8\pi G\varepsilon}{3c^2} + \frac{\Lambda c^2}{3} - \frac{kc^2}{a^2}\right), \quad (6)$$

$$\frac{\ddot{a}}{a} = -\frac{4\pi G}{3c^2}(\varepsilon + 3p) + \frac{\Lambda c^2}{3} + \frac{\dot{c}}{c}\frac{\dot{a}}{a} = -\frac{4\pi G}{3c^2}(\varepsilon + 3p) + \frac{\Lambda c^2}{3} + \frac{\dot{c}}{c}\frac{a}{\dot{a}}H^2. \quad (7)$$

Here $a$ is the cosmological expansion scale factor with its current value taken as 1. When we take time derivative of Equation (6), divid it by $2a\dot{a}$, and equate it with Equation (7) we get the general continuity equation:

$$\dot{\varepsilon} + 3\frac{\dot{a}}{a}(\varepsilon + p) = -\left[\left(\frac{\dot{G}}{G} - 4\frac{\dot{c}}{c}\right)\varepsilon + \frac{c^4}{8\pi G}\dot{\Lambda}\right]. \quad (8)$$

Equation (3) for the FLRW metric and perfect fluid energy-momentum tensor becomes:

$$\left[\left(\frac{\dot{G}}{G} - 4\frac{\dot{c}}{c}\right)\frac{8\pi G}{c^4}\varepsilon + \dot{\Lambda}\right] = 0, \quad (9)$$

therefore,

$$\dot{\varepsilon} + 3\frac{\dot{a}}{a}(\varepsilon + p) = 0. \quad (10)$$

Using the equation of state relation $p = w\varepsilon$, the solution for this equation is $\varepsilon = \varepsilon_0 a^{-3-3w}$, with $\varepsilon_0$ the current energy density of all the components of the universe. Here $a = a_0 = 1$, $w = 0$ for matter, and $w = 1/3$ for relativistic particles.

From the continuity equation (Equation 9), wen $\Lambda$ is constant, $\dot{G}/G = 4\dot{c}/c$. However, when $\Lambda$ is not constant, we are could write $\dot{G}/G = \sigma\dot{c}/c$ with $\sigma$ an unknown parameter. Then, from Equation (9), by defining $\varepsilon_\Lambda = c^4\Lambda/(8\pi G)$, wemay write

$$\frac{8\pi G}{c^4}\frac{\dot{c}}{c}(4-\sigma)\varepsilon = \dot{\Lambda}, \Rightarrow \frac{\dot{c}}{c} = \frac{c^4\Lambda}{8\pi G}\left(\frac{\dot{\Lambda}}{\Lambda}\right)\left(\frac{1}{(4-\sigma)\varepsilon}\right) \equiv \frac{\varepsilon_\Lambda}{(4-\sigma)\varepsilon}\frac{\dot{\Lambda}}{\Lambda},$$

$$\Rightarrow \varepsilon_\Lambda = \frac{\dot{c}}{c}\frac{\Lambda}{\dot{\Lambda}}(4-\sigma)\varepsilon \quad (11)$$

The unknown parameter $\sigma$ may be determined from the physics and by fitting the observational data. We have determined in the past [52] that $\sigma = 3$ analytically, i.e. $\dot{G}/G = 3\dot{c}/c$ and confirmed it by fitting the SNe1a data [37]. Thus, we must have $\varepsilon_\Lambda = \dot{c}\Lambda\varepsilon/(c\dot{\Lambda})$ from Equation (11).

Most commonly, one represents the variation of the constant through the scale factor powerlaw, e.g., $c = c_0 a^\alpha$, which results in $\dot{c}/c = \alpha\dot{a}/a = \alpha H$ where $\alpha$ is an unknown parameter. It results in very simple Freedmann equations. However, as $a \to 0$ the varying constant tends to zero or infinity depending on the sign of $\alpha$. Thus, it yields reasonable results only when $a = 1/(1+z)$ corresponding to relatively small redshift $z$. This led us to try another relation that resulted in:

$$c = c_0 f(a); G = G_0 f(a)^3; \text{ and } \Lambda = \Lambda_0 g(a), \quad (12)$$

with $f(a) = \exp(a^\alpha - 1), \quad (13)$

and $g(a)$ determined by substituting $\Lambda = \Lambda_0 g(a)$ in Equation (11) with $\sigma = 3$. The limitation of this form is that $c$ can decrease in the past *at most* by $1/e$ (= 1/2.7183), and $G$ can decrease by a factor of $e^{-3}$ (for positive $\alpha$ within the region of their applicability).

The first Friedmann equation, Equation (6), becomes

$$H^2 = \frac{8\pi G}{3c^2}\left(\varepsilon + \frac{\Lambda c^4}{8\pi G}\right) - \frac{kc^2}{a^2} = \frac{8\pi G}{3c^2}(\varepsilon + \varepsilon_\Lambda) - \frac{kc^2}{a^2}. \quad (14)$$

Here energy density $\varepsilon = \varepsilon_m + \varepsilon_r = \varepsilon_{m,0}a^{-3} + \varepsilon_{r,0}a^{-4}$ with subscript $m$ for matter and $r$ for radiation (relativistic particles, e. g., photons and neutrinos). Dividing by $H_0^2$, we get

$$\frac{H^2}{H_0^2} = \frac{8\pi G}{3c^2 H_0^2}\left(\varepsilon_{m,0}a^{-3} + \varepsilon_{r,0}a^{-4} + \varepsilon_\Lambda\right) - \frac{kc^2}{H_0^2 a^2}. \quad (15)$$

Therefore, for $a = 1$, i.e., at the current time

$$1 = \frac{8\pi G_0}{3c_0^2 H_0^2}\left(\varepsilon_{m,0} + \varepsilon_{r,0} + \varepsilon_{\Lambda,0}\right) - \frac{kc_0^2}{H_0^2}$$
$$\equiv \left(\Omega_{m,0} + \Omega_{r,0} + \Omega_{\Lambda,0}\right) - \frac{kc_0^2}{H_0^2}. \quad (16)$$

Here we have defined the current critical density as $\varepsilon_{c,0} = 3c_0^2 H_0^2/8\pi G_0$, $\Omega_{m,0} = \varepsilon_{m,0}/\varepsilon_{c,0}$, $\Omega_{r,0} = \frac{\varepsilon_{r,0}}{\varepsilon_{c,0}}$, and $\Omega_{\Lambda,0} = \varepsilon_{\Lambda,0}/\varepsilon_{c,0}$. Thus, by defining $\Omega_0 = (\Omega_{m,0} + \Omega_{r,0} + \Omega_{\Lambda,0})$, we may write Equation (16)

$$\Omega_{k,0} \equiv -\frac{kc_0^2}{H_0^2} = 1 - \Omega_0. \quad (17)$$

We must now express $\varepsilon_\Lambda$ using Equation (11) subject to the general constraint (Equation 9). This is somewhat convoluted. Nevertheless, following Gupta [38 – Appendix A], we may write

$$\frac{H^2}{H_0^2} = \exp(a^\alpha - 1)\left[\Omega_{m,0}a^{-3}\{1 + \alpha F(\alpha, a)\} + \Omega_{r,0}a^{-4} + \Omega_{\Lambda,0}\exp(a^\alpha - 1)\right] + \frac{1 - \Omega_{m,0} - \Omega_{r,0} - \Omega_{\Lambda,0}}{a^2}\exp[2(a^\alpha - 1)] \quad (18)$$

$$\equiv E(a)^2 \to E(z)^2 \text{ by substituting } a = 1/(1+z). \quad (19)$$

Here

$$F(\alpha, a) = a^3 \exp(a^\alpha - 1)\left[\int_1^a \frac{a'^{(\alpha-4)}}{\exp(a'^\alpha - 1)}da'\right]. \quad (20)$$

It is easy to see that $F(\alpha, 1) = 0 = F(\alpha, 0)$, and the equations reduce to the ΛCDM form when $\alpha = 0$, i.e., no variation of the constants:

$$\frac{H^2}{H_0^2} = (\Omega_{m,0}a^{-3} + \Omega_{r,0}a^{-4} + \Omega_{\Lambda,0}) - \frac{1-\Omega_{m,0}-\Omega_{r,0}-\Omega_{\Lambda,0}}{a^2}. \quad (21)$$

Let us now determine the proper distance $d_P$ between an observer and a source. We may write the FLRW metric in spherical spatial coordinates (Equation 4) as [51]

$$ds^2 = -c^2dt^2 + a(t)^2[dr^2 + S_k(r)^2(d\theta^2 + \sin^2\theta d\phi^2)]. \quad (22)$$

Here $S_k(r) = R\sin(r/R)$ for $k = +1$ (closed Universe); $S_k(r) = r$ for $k = 0$ (flat Universe), $S_k(r) = R\sinh(r/R)$ for $k = -1$ (open Universe), where $R$ is the parameter related to the curvature. The proper distance $d_P$ is determined at fixed time by following a spatial geodesic at constant $\theta$ and $\phi$. Then

$$ds = a(t)dr \Rightarrow d_P(t) = a(t)\int_0^r dr = a(t)r. \quad (23)$$

We could determine $r$ following a null geodesic from the time $t$ a photon is emitted by the source to the time $t_0$ it is detected by the observer with $ds = 0$ in Equation (22) at constant $\theta$ and $\phi$:

$$c^2dt^2 = a(t)^2dr^2 \Rightarrow \frac{cdt}{a(t)} = dr \Rightarrow r = \int_0^r dr = \int_t^{t_0}\frac{cdt}{a(t)}$$
$$\Rightarrow d_P = a(t_0)\int_t^{t_0}\frac{cdt}{a(t)}. \quad (24)$$

Now $dt = da \cdot dt/da = da/\dot{a} = da/a\dot{a}/a = da/aH$, and $a = 1/(1+z)$, $da = -dz/(1+z)^2 = -a^2 dz$. Therefore

$$dt = -\frac{adz}{H} = -\frac{adz}{\frac{H_0 H}{H_0}} = -\frac{adz}{H_0 E(a)}, \text{ and} \quad (25)$$

$$d_P = \frac{1}{H_0}\int_0^z \frac{cdz}{E(z)} = \frac{c_0}{H_0}\int_0^z \frac{\exp[((1+z)^{-\alpha}-1)]dz}{E(z)}. \quad (26)$$

Here $E(z)$ is given by Equations (18, 19).

Following Gupta [37], we may write the photons emitted by a source at the time $t_e$ are spread over a sphere of radius $S_k(d_P)$ and area $A_P(t_0)$ by the time photons reach the observer at the time $t_0$. The area of the sphere

$$A_P(t_0) = 4\pi S_k(d_P)^2. \quad (27)$$

The photon energy flux is defined as luminosity $L$ divided by the area in a stationary universe. When the Universe is expanding, then the flux is reduced by a factor $1 + z$ ($\equiv a^{-1}$) due to energy reduction of the photons from the change in their wavelength:

$$\lambda_0 a = \lambda_e \Rightarrow \lambda_0 = (1+z)\lambda_e.$$

The photon energy is thus altered by a factor of $1/(1+z)$ due to the expanding Universe.

We also need to determine how the increase in the time interval of the emitted photons affects the flux. The proper distance between two emitted photons separated by a time interval $\delta t_e$ is $c_e \delta t_e$ whereas the proper distance between the same two photons when detected by observation, is $(c_e \delta t_e)(1+z)$, and the time interval between the same two photons is $\delta t_0 = (c_e \delta t_e)(1+z)/c_0 = (1+z)f\delta t_e$ (since $c_e = c_0 f$, see Equation 12). Thus, the time duration between the photons has altered by a factor of $\delta t_e/\delta t_0 = 1/[(1+z)f]$, which is time dilation, that we have to consider in estimating the flux, and, therefore, in calculating the source distance.

The above two effects alter the photon energy flux cumulatively by a factor $1/[(1+z)^2 f]$. Thus, we may write the flux of photons energy $F_0$ received by an observer as

$$F_0 = \frac{L_{source}}{4\pi S_k^2(d_P)}(1+z)^{-2}f^{-1} \equiv \frac{L_{source}}{4\pi d_L^2}$$
$$\Rightarrow d_L = S_k(d_P)(1+z)\sqrt{f}. \quad (28)$$

Here $d_L$ is the luminosity distance of the source. The distance modulus $\mu$, derived from the measured absolute bolometric magnitude $M$ of the source at the luminosity distance $d_L$ and it's apparent magnitude $m$ if placed at a distance of 10 parsec, is by definition related to the luminosity distance $d_L$ [51]:

$$\mu \equiv m - M = 5\log_{10}\left(\frac{d_L}{1Mpc}\right) + 25 = 5\log_{10}\left(\frac{S_k(d_P)}{1Mpc}\right) +$$
$$5\log_{10}(1+z) + 2.5\log_{10}f(z) + 25. \quad (29)$$

We will also need to consider how the luminosity of source is itself altered due to the varying coupling constants in order to correlate the proper distance $d_P$ with the luminosity distance $d_L$. Here, by definition $f(z) = \exp[(1+z)^{-\alpha} - 1]$. We have determined $\alpha = 1.8$ [37, 52].

Let us see how to calculate $S_k(d_P)$ once we know $d_P$ from Equation (26). Recall $S_k(r) = R\sin(r/R)$ for $k = +1$ (closed Universe); $S_k(r) = r$ for $k = 0$ (flat Universe), $S_k(r) = R\sinh(r/R)$ for $k = -1$ (open Universe), where $R$ is the parameter related to the curvature and $k$ is in the units of $1/R^2$. Therefore, using Equation (17), we can write $R = (c_0/H_0)/\sqrt{|\Omega_0 - 1|}$, and using Equation (26), we can calculate $\mu$ (Equation 29).

***Varying coupling constants and their interdependence***: We have only considered how the variation of $c, G,$ and $\Lambda$ are related. Let us now see how $c$ and $G$ variations are related to the variation of the Planck constant $h$ and the Boltzmann constant $k$ [40].

Let us assume that the coupling constants evolve with the expansion of the Universe through scale factor $a$ as follows: $c = c_0 f_c(a)$; $G = G_0 f_G(a)$; $h = h_0 f_h(a)$; and $k_B = k_{B,0} f_k(a)$. Here subscript 0 on a coupling constant refers to its current value, and the subscript on the arbitrary function $f(a)$ identifies the associated coupling constant.

Consider now an exploding star of mass $M$ and radius $r$, such as a core-collapse supernova, where a fraction $\eta$ of the mass is converted through fusion ($\eta \sim 0.7\%$ for hydrogen to helium conversion) into the nuclear energy causing the explosion. Assume a fraction $\beta$ of the explosion energy is used up in countering the negative self-gravitational energy of the mass to bring it to zero, and the balance shows up as kinetic energy of the exploded particle (ignoring energy loss due to escaping neutrino and antineutrino particles). A fraction $\gamma$ of this kinetic energy thermalizes and is partially radiated away as photons. When distances are measured using the speed of light [36, 53], the evolution of the energies may be written

$$\eta M c^2 \times \beta = \frac{GM^2}{r} = \frac{GM^2}{r_c(c/c_0)} \Rightarrow \eta \beta c^3 = \frac{GMc_0}{r_c}, \tag{30}$$

where $r_c$ is the stellar radius independent of the speed of light (similar to the commoving distance in cosmology) defined by $r \equiv r_c(c/c_0)$. Thus,

$$\eta \beta c_0^3 f_c(a)^3 = \frac{G_0 f_G(a) M c_0}{r_c}. \tag{31}$$

Local energy conservation over each slice of the cosmic time, i.e., scale factor $a$, leads to $f_G(a) = f_c(a)^3$, i.e., $G \sim c^3$.

Now consider the thermalized kinetic energy of $N$ particles, comprising mass $M$, at temperature $T$. Then

$$\eta(1-\beta) M c^2 \times \gamma = N k_B T. \tag{32}$$

This means that $c^2 \sim k_B T$. Since $T$ is an arbitrary measure of thermal energy, $k_B \sim c^2$, i.e., $f_k(a) = f_c(a)^2$.

Finally, consider that a fraction $\delta$ of the thermal energy generates $N_\lambda$ number of photons of wavelength $\lambda$. Then

$$\delta N k_B T = \frac{N_\lambda h c}{\lambda}. \tag{33}$$

Since $N$ and $N_\lambda$ are conserved in an evolutionary (expanding) universe, and we know $\lambda \sim a$, we must have $k_B \sim hc$. But $k_B \sim c^2$, which leads to $h \sim c$, i.e., $f_h(a) = f_c(a)$.
In summary,

$$f_c(a) = f_h(a) = f_k(a)^{1/2} = f_G(a)^{1/3} \equiv f(a), \text{ or}$$

$$c = c_0 f(a), G(a) = G_0 f(a)^3, h(a) = h_0 f(a),$$
$$\text{and } k_B = k_{B,0} f(a)^2. \tag{34}$$

If $\eta, \beta, \gamma,$ and $\delta$ are functions of the scale factor $a$, then they can be absorbed in functions representing the variations of the constants without affecting our findings.

The above analysis has general applicability and is not confined to supernovae explosions. The supernova explosion was chosen as it involves all four types of energy conversions needed for the analysis. Also, it is easy to see that including the energy loss from escaping neutrinos and antineutrinos does not affect our study.

## 3. SUPERNOVAE HUBBLE DIAGRAM

Our focus in this section will be to explore how well the Pantheon supernovae type Ia (SNeIa) data [45] fit the CCC model. SNeIa are the best standard candles for measuring cosmological distances, and thus this fit is essential for a model to qualify for further studies.

*Luminosity evolution of SNeIa in the CCC model*: The peak SNIa luminosity, $L_{SN}$, is proportional to the mass of the nickel synthesized in the white dwarf explosion resulting in the SNIa, which is proportional to Chandrasekhar mass $M_{Ch}$ of the exploding star [15, 54, 55]. The explosion energy is partially used up to counter the gravitational binding energy $E_{gr}$ of the star, and the balance is converted into the kinetic energy $E_{ke}$. A fraction of this kinetic energy is radiated out and observed as SNIa luminosity. Therefore,

$$\eta M_{Ch} c^2 \equiv E_{kin} + E_{gr}. \tag{35}$$

Here, $\eta$ represents the efficiency of mass-to-energy conversion. Now, the Chandrasekhar mass and the radius $r_{wd}$ of a white dwarf are given by [56],

$$M_{ch} = 0.21 \left(\frac{Z}{A}\right)^2 \left(\frac{hc}{Gm_p^2}\right)^{3/2} m_p, \text{ and} \tag{36}$$

$$r_{wd} \approx \frac{h^2}{10 m_e m_p^{5/3} G} \left(\frac{Z}{A}\right)^{5/3} M_{ch}^{-1/3}. \tag{37}$$

Here $Z$ is the atomic number, $A$ is the atomic mass number, $m_p$ is the proton mass, and $m_e$ is the electron mass. The gravitational binding energy is [e.g., 56]

$$E_{gr} \approx -\frac{GM_{ch}^2}{r_{wd}}. \tag{38}$$

We could now use Equation (34) to see how the above quantities vary with the expansion of the Universe. It should be noted that the function $f(a)$ truly represents the variation of the dimensionless ratio of a constant with its currently measured value. Thus, we tacitly refer to the variation of this dimensionless ratio when we say the variation of a constant. It is worth noting that the quantity $hc/Gm_p^2$ in equation (36) defining Chandrasekhar mass is dimensionless but scales as $f^{-1}$ (assuming $m_p$ is either constant or varies at a rate negligible compared to the variation of $c, h,$ and $G$). It plays an essential role in this study.

We may now write the following scaling relations for the white dwarf of Chandrasekhar mass $M_{Ch}$:

$$M_{Ch} c^2 \sim f^{1/2}, \tag{39}$$

$$r_{wd} \sim f^{-1/2}, \text{ and} \tag{40}$$

$$E_{gr} \sim \frac{GM_{Ch}^2}{r_{wd}} \sim f^{1/2}. \tag{41}$$

Now that $M_{Ch}c^2$ and $E_{gr}$ are both scaling as $f^{1/2}$ in the equation, we may write the scaling of the energy $E_{SN}$ contributing to the SN Ia luminosity

$$E_{SN} \propto E_{kin} = (\eta M_{Ch}c^2 - E_{gr}) \sim f^{1/2}. \tag{42}$$

We also need to consider how the energy levels of electrons involved in the emission of radiation, which affects the luminosity, scales when the coupling constants vary. The energy levels are proportional to the Rydberg unit of energy: $R_y = hcR_\infty$ where $R_\infty = m_e e^4/8\epsilon_0^2 h^3 c$ is the Rydberg constant. Here $e$ is the electron charge and $\epsilon_0$ is the permittivity of space given by $c = 1/\sqrt{\mu_0 \epsilon_0}$ with $\mu_0 = 4\pi \times 10^{-7}$ H m$^{-1}$ the permeability of free space. All masses, electric charges, and permeability are constant in the CCC model. This means,

$$R_\infty \sim c^4/(h^3 c) \sim f^0, \text{ and} \tag{43}$$

$$R_y \sim f^2 \times f^0 \sim f^2. \tag{44}$$

Therefore, the number of photons $N$ released in the SN Ia explosion

$$N \propto E_{SN}/R_y \sim f^{-3/2} \Rightarrow N = N_0 f^{-3/2} \Rightarrow N_0 = N f^{3/2}. \tag{45}$$

Let us now consider the evolution of the photon energy itself. It is given by $ch/\lambda$ where $\lambda$ is the photon wavelength. The lower energy of a photon at the time of emission by a factor $f^2$ due to the lower values of $c$ and $h$ is offset by the increase in the photon energy at the time of its detection due to the higher value of $c$ and $h$ by a factor $f^{-2}$ for a given photon wavelength $\lambda$. However, the photon wavelength does expand due to the expansion of the Universe as the scale factor $a$, and this must be taken into account in calculating the detected photon energy and luminosity.

It is now apparent that we must modify Equations (28) and (29) to take into account Equation (45).

$$F_0 = \frac{L_{source}}{4\pi S_k^2(d_P)}(1+z)^{-2}f^{-1}f^{3/2} \equiv \frac{L_{source}}{4\pi d_L^2}$$
$$\Rightarrow d_L = S_k(d_P)(1+z)f^{-1/4}; \tag{46}$$

$$\mu = 5 \log_{10}\left(\frac{S_k(d_P)}{1Mpc}\right) + 5 \log_{10}(1+z) + 25 - 1.25 \log_{10} f(z). \tag{47}$$

Other factors could also affect the SNeIa luminosity as a standard candle. One such factor is the dependence of SNeIa luminosities on the metallicities of their host galaxies [e.g., 57]. They showed that the SNeIa in high metallicities hosts are $0.14 \pm 0.10$ magnitude brighter than those in the low metallicity hosts. If we consider that the metallicities of the galaxies increase with their age, i.e., they increase with increasing scale factor $a$, and that the calibration of the SNIa standard candle luminosity is done from observations in galaxies with $a \approx 1$, we may write the magnitude decrease with $a$ as $0.14(1-a)$. This corresponds to the magnitude decreasing with increasing redshift $z$ as $0.14z/(1+z)$. Equation (48) should therefore be modified as follows:

$$\mu = 5 \log_{10}\left(\frac{S_k(d_P)}{1Mpc}\right) + 5 \log_{10}(1+z) + 25$$
$$- 1.25 \log_{10}[\exp\{(1+z)^{-\alpha} - 1\}] - 0.14\left(\frac{z}{1+z}\right). \tag{48}$$

We have expressed the form of function $f(z) = \exp\{(1+z)^{-\alpha} - 1\}$ explicitly in Equation (48). First three terms are the same as in the ΛCDM model. In a flat universe $S_k(d_P) = d_P$. The $\alpha$ parameter may be considered as representing the strength of the variation of the constants. We have found $\alpha = 1.8$ analytically [52] and confirmed it from the analysis of various observations [34, 35, 36, 37, 38, 39, 43]. We showed graphically the dependence of key cosmological parameters $H_0$, $\Omega_{m,0}$, $\Omega_{\Lambda,0}$, and $\Omega_{k,0}$ on $\alpha$ in Fig. 1 of reference [38]. It may be noted that irrespective of the value of $\alpha$, $f(z) \to 1/e$ as $z \to \infty$ such as for the surface of last scattering for cosmic microwave background and for the big-bang nucleosynthesis.

***Results***: We use Pantheon SNeIa data (Scolnic et al. 2018) for determining parameter for the CCC and the ΛCDM models. The redshift range of the Pantheon Sample is $0.01 < z < 2.3$. The data has observation in terms of the apparent magnitude, and we added 19.35 to obtain normal distance modulus numbers as suggested by Scolnic (private communication). We used the Matlab curve fitting tool to fit the data by minimizing $\chi^2$. Here $\chi^2$ is the weighted summed square of residual of $\mu$

$$\chi^2 = \sum_{i=1}^{N} w_i \left[\mu(z_i; H_0, p_1, p_2 \ldots) - \mu_{obs,i}\right]^2,$$

where $N$ is the number of data points, $w_i$ is the weight of the $i$th data point $\mu_{obs,i}$ determined from the measurement error $\sigma_{\mu_{Obs,i}}$ in the observed distance modulus $\mu_{obs,i}$ using the relation $w_i = 1/\sigma_{\mu_{Obs,i}}^2$, and $\mu(z_i; H_0, p_1, p_2..)$ is the model calculated distance modulus dependent on parameters $H_0$ and all other model-dependent parameters $p_1, p_2$, etc. For the most general case $p_1 \equiv \Omega_{m,0}, p_2 \equiv \Omega_{\Lambda,0}$, and $p_3 \equiv \alpha$ (since for the matter dominated epoch $\Omega_{r,0}$ could be ignored). Thus, we have four free parameters to consider in fitting the data: three when fitting a ΛCDM model and four when fitting the CCC model. The extra parameter for the CCC case determines the strength of the variation of the evolving constants. In fact, the ΛCDM model can be considered a special case of the CCC model when $\alpha = 0$.

We first attempted to keep all four parameters free and determine the value of $\alpha$ that yields the minimum $\chi^2$. We expected $\alpha$ to come out zero if the ΛCDM model is the best model to fit the Pantheon sample. But it did not. This outcome prompted us to explore the sensitivity of $\chi^2$ against the variation of $\alpha$. In fact, $\chi^2$ was stable against $\alpha$ variation within certain limits. This meant the SNeIa data could not determine a unique value of $\alpha$, which must be determined from other observations as was done in an earlier paper [38] wherein we concluded $\alpha = 1.8$. We will, therefore, use $\alpha = 1.8$ everywhere in this paper.

We fitted the Pantheon data for the CCC model using Equation (48) and determined the cosmological parameters: the Hubble constant $H_0 = 70.83 \, (\pm 0.66)$ km s$^{-1}$ Mpc$^{-1}$, the matter density $\Omega_{m,0} = 0.2708 \, (\pm 0.0626)$, and the dark energy density $\Omega_{\Lambda,0} = 0.1754 \, (\pm 0.1727)$. The numbers in parenthesis indicate 95% confidence level. The last two numbers indicate that the Universe is negatively curved. For fitting the data to the benchmark ΛCDM model, we drop the last two terms in Equation (48), assumed and $S_k(d_P) = d_P$ for flat Universe, and substituted $\Omega_{m,0} = 0.3$ and $\Omega_{\Lambda,0} = 0.7$. As expected, the fit resulted in Hubble constant $H_0 = 69.96 \, (\pm 0.26)$ km s$^{-1}$ Mpc$^{-1}$. We immediately notice that the matter energy density $\Omega_{m,0}$ determined using the CCC model is the same as for the benchmark model within the 95% confidence level, whereas the dark energy density $\Omega_{\Lambda,0}$ is vastly different, meaning that the latter trades with curvature energy density. The two fits are presented in Figure 2.

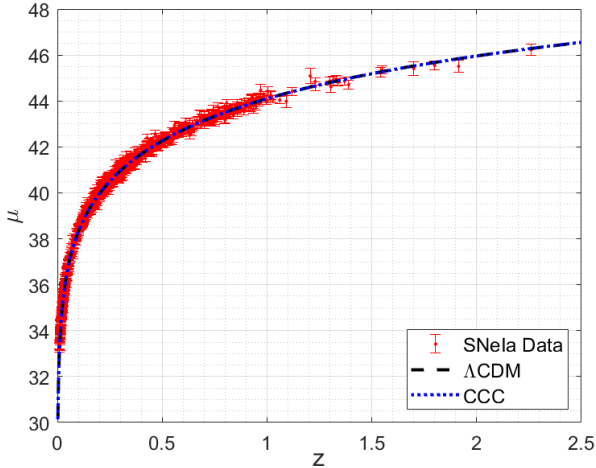

**FIGURE 2**
Pantheon sample supernovae type Ia data fit using the benchmark ΛCDM model and the covarying coupling constant (CCC) model. The two fits are graphically indistinguishable.

As is evident, the fitted data curves for the two models are graphically indistinguishable. Even the $\chi^2$ per degree of freedom values (0.9904 for the ΛCDM model and 0.9876 for the CCC model) are not significantly different. Thus, one may conclude that the two cosmological models are equally good. The CCC model must be tested against other cosmological observables. One additional conclusion would be that the coupling constants variability function $f(z) = \exp\{(1 + z)^{-1.8} - 1\}$ used for fitting the data can be employed for constraining the coupling constants.

## 4. QUASARS HUBBLE DIAGRAM

The 'extreme Population A (xA) quasars' approaching, sometimes exceeding the Eddington limit, are a class of quasars that could serve as standard candles to measure distances too large for supernovae type Ia to be observable [e.g., 46, 58, 59, 60, 61, 62, 63, 64].

Three conditions need to be satisfied for the possible use of xA quasars as standard candles [64]:
1. Eddington ratio $L/L_E \equiv \lambda_E$ is constant where $L$ is the luminosity of the quasar and $L_E$ is its Eddington luminosity.
2. The black hole mass can be expressed with the virial relation $M = r\delta v^2/G$. Here $r$ is the radius of the broad-line region (BLR) of the emitted radiation, $\delta v$ is the virial velocity in the region.
3. They should have spectral invariance, i.e., the ionization parameter $U = Q(H)/(4\pi r^2 n_H c)$ should be constant. Here $Q(H)$ is the number of hydrogen ionizing photons, $n_H$ is the hydrogen number density, and $c$ is the speed of light.

We will study if the above conditions are affected under the covarying coupling constants scenario.

The luminosity $L$ of a quasar with a black hole of mass $M$ that is accreting mass at a rate $\dot{M}$ from a disk of the inner diameter $r_{in}$ and outer diameter $r_{out}$ with $r = r_{in} \ll r_{out}$ is given by [56]

$$L = \frac{1}{2}\left(\frac{GM\dot{M}}{r_{in}}\right), \tag{49}$$

whereas the Eddington luminosity $L_E$ is

$$L_E = \frac{4\pi c G M m_p}{\sigma_T}. \tag{50}$$

Here $m_p$ is the proton mass and $\sigma_T$ is the Thomson scattering cross-section representing the scattering of photons by electrons. Since $L_E$ is the limiting luminosity, $L \leq L_E$. Therefore, from equations (49) and (50)

$$\dot{M} \leq 8\pi r c m_p / \sigma_T. \tag{51}$$

Let us now see how the above three expressions scale when the coupling constants vary as in Equation (34), i.e.,

$$G \sim c^3 \sim h^3 \sim k^{3/2} \sim f^3, \text{ and } \frac{\dot{G}}{G} = 3\frac{\dot{c}}{c} = 3\frac{\dot{h}}{h} = \frac{3}{2}\frac{\dot{k}}{k}. \tag{52}$$

These relations are independent of the form of $f(a)$, i.e., $f(z)$. However, the form we have successfully used for SNeIa data fit above is

$$f = \exp(a^\alpha - 1) \equiv \exp[(1 + z)^{-\alpha} - 1]. \tag{53}$$

Here $\alpha$ is the parameter representing the strength of the variation of the constants. We have found $\alpha = 1.8$ analytically

[52] and confirmed it from the analysis of various observations [34, 35, 36, 37, 38, 39, 43].

We may now write the scaling of the expressions (Equations 49 – 51), realizing that any distance is measured using the speed of light, i.e., $r \sim c \sim f$. However, we should first see how the Thomson scattering cross-section $\sigma_T$ scales. It is given by

$$\sigma_T = \frac{8\pi}{3}\left(\frac{q^2}{4\pi\epsilon_0 mc^2}\right)^2. \qquad (54)$$

Here $q$ is the elementary charge, and $m$ is the mass of the particle. As discussed above, $\epsilon_0$ is the permittivity of space that is related to the speed of light through $c = 1/\sqrt{\epsilon_0 \mu_0}$ where $\mu_0 = 4\pi \times 10^7$ H m$^{-1}$ is the permeability of space. Thus, $\epsilon_0 \propto 1/c^2$. Since charge and mass are considered invariant when coupling constants vary, we get $\sigma_T \sim f^0$. Therefore,

$$\dot{M} \leq 8\pi r c m_p / \sigma_T \sim f^2, \qquad (55)$$

$$L = \frac{1}{2}\left(\frac{GM\dot{M}}{r_{in}}\right) \sim \frac{f^3 \dot{M}}{f} \sim f^4, \qquad (56)$$

$$L_E = \frac{4\pi cGMm_p}{\sigma_T} \sim f^4. \qquad (57)$$

Thus, the Eddington ratio, $L/L_E \sim f^0$, is unaffected by the variation of the coupling constants: One cannot use it for constraining the variation of the constants.

Let us now consider the black hole mass determination. Again, we may write

$$M = r\delta v^2/G. \qquad (58)$$

Since $r \sim f$, and velocity is the time derivative of length measured with the speed of light $c$, it also scales as $f$. Therefore, the mass of the black hole, $M \sim ff^2/f^3 \sim g^0$, is unaffected by the constants' variation: We cannot consider it for constraining the constants.

Next, we should examine the spectral invariance, which may be rewritten as

$$U = \frac{Q(H)}{4\pi r^2 n_H c}. \qquad (59)$$

The hydrogen number density is inversely proportional to volume, i.e., $n_H \propto 1/length^3 \sim f^{-3}$. Therefore, $\sim Q(H)/(f^2 f^{-3} f) \sim Q(H) g^0$. But $Q(H)$ represents the number of hydrogen ionizing photons, which does not change with $z$ under the CCC scenario. While the ionizing photon energy evolves with $z$, it is exactly offset by the evolution of the energy required to ionize hydrogen atoms [38] and is discussed above in the paragraph under Equation (45). Thus, $U \sim f^0$, i.e., $U$ does not vary on account of the varying coupling constants and, therefore, cannot be used for constraining them.

***Luminosity evolution of xA quasars in the CCC model***: We will use the same CCC approach to fit the quasar data that we used to fit the SNeIa data above (see also [38]). However, the approach will be modified since, among other things, quasar luminosities depend strongly on their black hole masses that are known to increase with the redshift [e.g., 47].

From Equation (56), we may write the source luminosity $L_{source}$ in the CCC universe in terms of the standard luminosity $L_s$ as $L_{source} = L_s f(z)^4$, i.e., $L_s = L_{source} f(z)^{-4}$. Therefore, Equation (28) is modified to

$$F_0 = \frac{L_{source}}{4\pi S_k^2(d_P)}(1+z)^{-2} f^{-5} \equiv \frac{L_{source}}{4\pi d_L^2}$$
$$\Rightarrow d_L = S_k(d_P)(1+z)f^{5/2}. \qquad (60)$$

Since the mass of a quasar increases with $z$ [e.g., 47], its luminosity (Equation 49) also increases with $z$. Typically, such increases are expressed using power law. However, any suitable function can be used. Let it be $g(z)$: $M_z = M_{z,0} g(z)^q$. Since the luminosity is directly proportional to $M$, the net effect is to correct the flux by an additional factor of $g(z)^{-q}$. This mass dependence of quasars' luminosity is already included in determining the luminosity using the standard model background [e.g., 46, 47, 64). This mass dependence must therefore be taken into consideration for the CCC model as well. We may, therefore, write the luminosity distance $d_L$

$$d_L = S_k(d_P)(1+z)f(z)^{5/2} g(z)^{q/2}, \qquad (61)$$

and the distance modulus $\mu$

$$\mu \equiv 5\log\left(\frac{d_L}{1\text{Mpc}}\right) + 5\log(1+z) + 25 = 5\log\left(\frac{S_k(d_P)}{1\text{Mpc}}\right) + 25 + 12.5\log f(z) + 2.5q\log g(z). \qquad (62)$$

***Results***: *Let* us consider the Hubble diagram of the xA quasars as studied by Mariziani and collaborators (e.g., Mariziani and Sultenic 2014, Dultzin et al. 2020, Figure 3). With $d_P$ from equation (26) substituted in equation (62), we can fit the quasar data [64], which has 253 data points, using the CCC cosmological parameter determined with SNeIa data fit. However, the parameters in the last term in Equation (62) have to be determined by fitting the data. These parameters can then be checked against the same parameters obtained by fitting the quasar mass vs. redshift data (e.g., Vestergaard & Osmer 2009). The $z - \mu$ fit can then be compared with the fit with the ΛCDM model ($\alpha = 0, q = 0$).

The most convenient form of the $g$ function is the same as for the $f$ function. Nevertheless, we will try other functions as well, i.e., the powerlaw and $g$ with the same form as for $f$, i.e., $g = \exp[(1+z)^{-\beta} - 1]$, but not constraining $\beta = 1.8$ as we normally do for $f$. We find the latter to be the best among those we have tried, as discussed later in this paper, and that is what we have used in Equation (62) to fit the 253 xA quasars data points.

The fit results for the ΛCDM model with $H_0 = 70$ km s$^{-1}$ Mpc$^{-1}$, $\Omega_{m,0} = 0.3$, and $\Omega_{\Lambda,0} = 0.7$, and for the CCC model with $H_0 = 70.83$ km s$^{-1}$ Mpc$^{-1}$, $\Omega_{m,0} = 0.2708$, and $\Omega_{\Lambda,0} = 0.1754$ are shown in Figure 3. The fit to the 253 data points for the ΛCDM model yields $\chi^2 = 456.0$, whereas for the CCC model, the fit gives $\chi^2 = 421.1$. The $\chi^2$ per degree of freedom values are 1.802 for the ΛCDM model and 1.678 for the CCC model. Thus, the CCC model appears to yield a significantly better fit, but it is partly due to the two unconstrained parameters $q$ and $\beta$ in the CCC model. The fit determined these parameters as $q = -7.462_{-8.885}^{-6.038}$ and $\beta = 1.046_{0.7196}^{1.3730}$ which are about the same as determined below, considering their 95% confidence bounds as shown by superscripts and subscripts of the values of respective parameters.

*Quasar mass evolution*: We have taken the data of the large bright quasar survey (LBQS) included in the Vestergaard and Osmer paper [47] to derive parameters of a powerlaw function and two $f$ type functions for $g$ function. LBQS data is for 978 quasars, with their redshift ranging from $z = 0.202$ to $3.364$ and mass ranging from $M_Q = 10^{7.17} M_\odot$ to $10^{10.53} M_\odot$ provided on $\log M_Q$ scale. It covers the range of data we will use for the Hubble diagram from Dultzin et al. [64]. The results are presented in Table 1. Figure 3 shows the data fit corresponding to the best fit function among the three in Table 1. The relevant parameters in the table are $q$ and $\beta$ as they relate to the $g$ function; $M_{Q,0}$ is irrelevant for scaling purposes as it is implicitly included in the first term of Equation (16).

Comparing the $q$ and $\beta$ values corresponding to the minimum $\chi^2$ (= 112.9, i.e., the middle $g$ function row) in the table with those determined from the $z - \mu$ fit above, we see that they are close and well within the 95% confidence bounds of their respective values. This match establishes that the CCC cosmology is a viable alternative to the ΛCDM cosmology in the context of the present work.

The evolution of a quasar's mass should thus be included in calibrating its observed luminosity for using it as a standard candle.

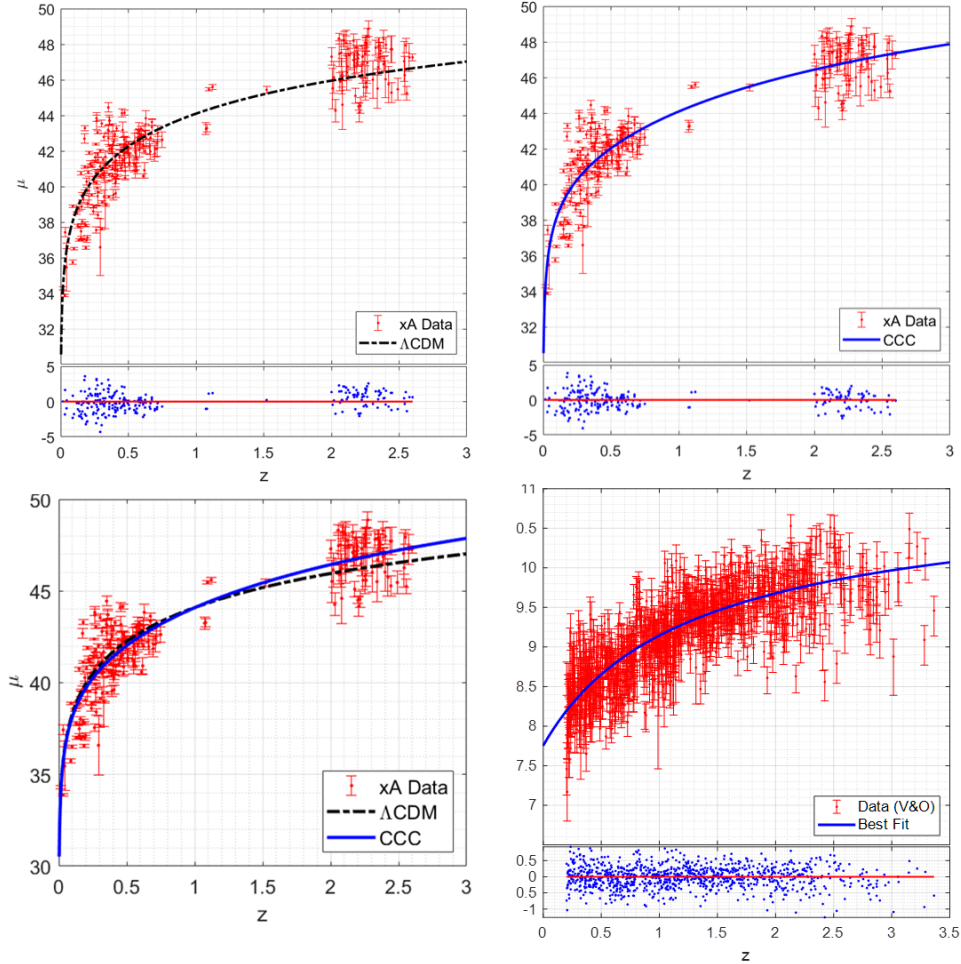

**FIGURE 3**
Hubble diagram of xA quasars fitted with the CCC model and the benchmark ΛCDM model and their mass function. Top-left: The data fit and the residuals plot using the CCC model. Top-right: The data fit and the residuals plot using the ΛCDM model. Bottom-left: Comparison of the two models fitted to the data [46]. Bottom-right: Mass variation of the quasars with redshift [47]. The residual plots are shown as well at the bottom of single model fits.

Table 1 The mass function parameters determined by fitting data with a powerlaw and alternative $g(z)$ functions. The values in subscripts and superscripts represent 95% confidence bounds.

| Function $M_Q = M_{Q,0} g^q$ | $\log M_{Q,0}$ | $q$ | $\beta$ | $\chi^2$ | $R^2$ |
|---|---|---|---|---|---|
| $g = z$, i.e., $\log M_Q = \log M_{Q,0} + q \log z$ | $9.162^{9.183}_{9.140}$ | $0.6863^{0.7188}_{0.6539}$ | NA | 114.7 | 0.638 |
| $g = \exp[(1+z)^{-\beta} - 1]$, i.e., $\log M_Q = \log M_{Q,0} + \dfrac{q}{2.3026}[(1+z)^{-\beta} - 1]$ | $7.735^{7.890}_{7.580}$ | $-7.33^{-6.149}_{-8.510}$ | $0.8424^{1.149}_{0.5361}$ | 112.9 | 0.6439 |
| $g = \exp[(1+z)^{-\beta} - 1]$ with $\beta = 1.8$ | $7.261^{7.356}_{7.167}$ | $-6.289^{-5.986}_{-6.591}$ | 1.8 fixed | 117.3 | 0.63 |

## 5. GAMMA-RAY BURSTS HUBBLE DIAGRAM

Gamma-ray bursts are the most energetic explosions in the Universe and thus can be observed in the galaxies formed at the earliest epochs. The gamma radiation from the bursts lasts for a very short duration (fraction of a second to a few minutes) followed by a longer-lasting after-glow observable at X-ray to radio wavelengths, the latter over several weeks. The energy released in a typical gamma-ray burst is of the order of the energy radiated by the Sun over its lifetime. GRBs' extremely high luminosity makes them observable at redshifts up to $z\sim10$ [65], corresponding to the time when the Universe was only about half a billion years in age. They occur at the rate of about one a day in the observable universe.

Unlike SNeIa, the light curves of GRBs are highly varied, and it is difficult to determine their luminosity from their observed data. Many schemes have been developed to categorize GRBs and estimate their luminosities in order to use them as standard candles. Several empirical correlations using parameters of the light curves and spectra have been proposed to standardize the GRBs' luminosity. They include correlations between (i) the GRB spectrum lag time $\tau_{lag}$ and isotropic peak luminosity $L$ ($\tau_{lag} - L$ relation) [66]; (ii) the peak energy $E_p$ of the $\nu F_\nu$ spectrum and the isotropic equivalent energy $E_{iso}$ ($E_p - E_{iso}$ relation) [67]; (iii) the time variability $V$ and isotropic peak luminosity ($V - L$ relation) [68], (iv) the peak energy of the $\nu F_\nu$ spectrum and the isotropic peak luminosity ($E_p - L$ relation) [69]; (v) the minimum rise time $\tau_{RT}$ of the light curve and isotropic peak luminosity ($\tau_{RT} - L$ relation) [70]; (vi) the peak energy and collimation-corrected energy $E_\gamma$ ($E_p - E_\gamma$ relation) [71, 72]; (vii) the peak luminosity, the time at the end of the plateau emission phase $\tau_a$, and the luminosity at the end of the plateau phase $L_a$ ($L - \tau_a - L_a$ relation - Dainotti 3D Fundamental Plane) [73, 74, 75]; and some more [e.g., 76, 77, 78, 79, 80, 81, 82, 83].

A cosmological model is typically used to standardize GRB correlations, which leads to the *circularity* problem (i.e., using a model to calibrate data and use the data to test the model). Several methods to avoid this problem are discussed by Liu et al. [84]. They used the Amati correlation that connects the spectral peak energy $E_P$ and isotropic equivalent radiated energy $E_{iso}$ [85] and improved it by employing the Gaussian copula statistical tool on Amati correlation. They thus obtained a model-independent Hubble diagram of A220 GRB samples [86] using Pantheon SNeIa data [45] to calibrate the correlation. The calibrated GRB data can be used to constrain a cosmological model. It has 220 $\mu - z$ data points.

Escamilla-Rivera et al. [48, 49] developed a new model-independent method to calibrate GRBs using SNeIa data [45] and cosmography. They used machine learning architecture by combining a *Recurrent Neural Network* and a *Bayesian Neural Network*. The machine learning comprises a deep learning architecture capable of producing a trained homogeneous sample of GRBs luminosity distance. This approach, combined with the GRbs' observed redshift, yields their Hubble diagram, i.e., $\mu - z$ data (141 points).

***Luminosity evolution of GRBs in the CCC model***: Let us consider Equation (60) for xA quasars. It includes the scaling of quasars luminosity by $f^{-4}$ (Equation 56) due to accreting mass on black holes. This is not relevant for GRBs. Thus, for the flux received, Equation (60) becomes

$$F_0 = \frac{L_{source}}{4\pi S_k^2(d_P)}(1+z)^{-2} f^{-1} \equiv \frac{L_{source}}{4\pi d_L^2}$$
$$\Rightarrow d_L = S_k(d_P)(1+z)f^{1/2}. \quad (63)$$

Just like in the cases of supernovae and quasars, we should expect an evolution of GRBs mass with cosmic time, i.e., with the redshift $z$. Let us write it as $M_z = M_{z,0} g(z)^q$ with $g(z)$ being the function representing the GRBs mass evolution. Assuming the luminosity of a GRB is directly proportional to its mass $M$, the net effect is to correct the flux by an additional factor of $g(z)^{-q}$. This mass dependence must therefore be taken into consideration for the CCC model applied to GRBs. We may, therefore, write the luminosity distance $d_L$

$$d_L = S_k(d_P)(1+z)f(z)^{1/2} g(z)^{q/2}, \quad (64)$$

and the distance modulus $\mu$

$$\mu \equiv 5\log\left(\frac{d_L}{1\text{Mpc}}\right) + 5\log(1+z) + 25 = 5\log\left(\frac{S_k(d_P)}{1\text{Mpc}}\right) +$$
$$5\log(1+z) + 25 + 2.5\log f(z) + 2.5 q \log g(z). \quad (65)$$

***Results***: Let us consider the Hubble diagram of the GRBs quasars as calibrated by Escamilla-Rivera et al. [49]. We will follow the same procedure we developed for fitting the quasars

Hubble diagram. The fit results for the ΛCDM model with $H_0 = 70$ km s$^{-1}$ Mpc$^{-1}$, $\Omega_{m,0} = 0.3$, and $\Omega_{\Lambda,0} = 0.7$, and for the CCC model with $H_0 = 70.83$ km s$^{-1}$ Mpc$^{-1}$, $\Omega_{m,0} = 0.2708$, and $\Omega_{\Lambda,0} = 0.1754$ are shown in Figure 4. The fit using the CCC model requires determining $q$ and $\beta$ (recall $g = \exp[(1+z)^{-\beta} - 1]$). We found a strong degeneracy between $q$ and $\beta$, and therefore decided to eliminate $q$ by setting it to unity. The data fit then yielded $\beta = 0.3315\ (\pm 0.0865)$ without compromising the goodness of fit.

The fit to the 141 data points for the ΛCDM model yields the goodness of fit parameters for 141 degrees of freedom (DOF) as $\chi^2 = 203.9$, i.e., $\chi^2/\text{DOF} = 1.4461$. The CCC model data fit with DOF = 140 gives $\chi^2 = 131.5$, i.e., $\chi^2/\text{DOF} = 0.9393$. Although the CCC model appears to yield a significantly better fit, it requires a mass function that affects the GRBs luminosity, i.e., $g(z) = \exp[(1+z)^{-\beta} - 1]$ in Equation (65). Nevertheless, we believe that the CCC model provides a significantly better fit over the full range of the redshift than does the ΛCDM model. Is CCC a better model for using GRBs as standard candles?

Figure 4 also includes a variation of the standard ΛCDM model wherein we retained the last term in Equation (65) with $q = 1$ and determined $\beta = 1.252\ (\pm 0.600)$. We dubbed this model ΛCDM*. This yielded $\chi^2 = 136.3$ and $\chi^2/\text{DOF} = 0.9736$, much better than for the standard ΛCDM model. However, fits for both the models are unsatisfactory: ΛCDM at low redshifts (evident from the residuals plot) and ΛCDM* at high redshift ($z > 6$).

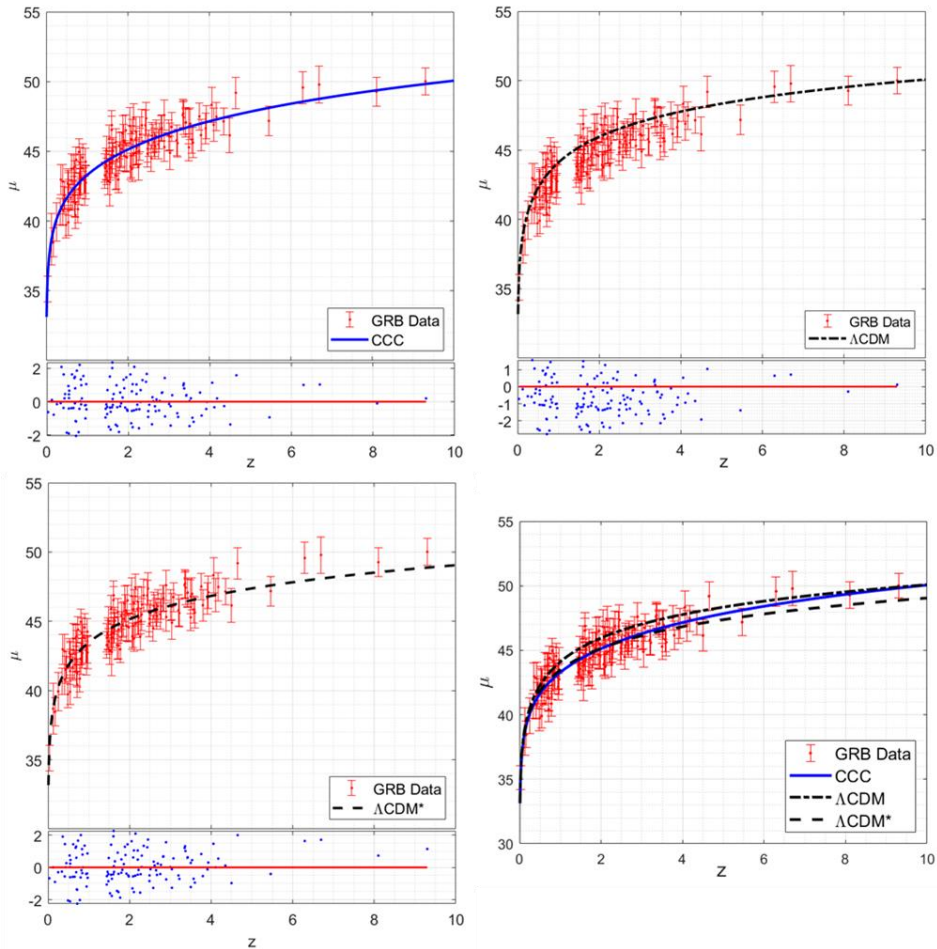

**FIGURE 4**
Gamma-ray burst data fit using the CCC model and the benchmark ΛCDM model. Top-left: The CCC model data-fit and the residuals plot. Top-right: The ΛCDM model data-fit and the residuals plot. Bottom-left: The ΛCDM* model data-fit and the residuals plot. Bottom-right: Comparison of the three models fitted to the data Escamilla-Rivera et al. [49].

With $q = 1$ and $\beta$ approximated as 1/3, the CCC model fit can thus be considered to make a prediction that the mass of GRBs evolves as

$$\log_{10}\left(\frac{M_{grb}}{M_{grb,0}}\right) = \log_{10}(\exp[(1+z)^{-\beta} - 1])$$
$$= ((1+z)^{1/3} - 1)/\ln(10), \qquad (66)$$

or simply as

$$\ln(M_{grb}/M_{grb,0}) = ((1+z)^{1/3} - 1). \qquad (67)$$

We have plotted the mass evolution of GRBs against their redshift in Figure 5, along with the same for quasars and SNeIa. It is evident that the GRB mass increases steadily with $z$, almost linearly, whereas the other two increase rapidly at low $z$ but tends to flatten out at higher $z$. We suggest that such mass variation and related luminosity evolution should be taken into consideration when calibrating GRB data.

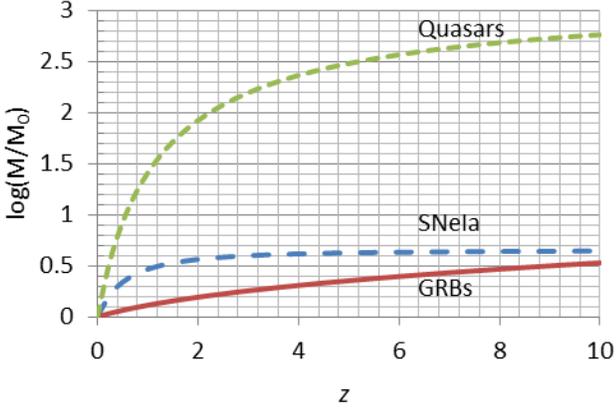

**FIGURE 5**
Mass evolution with redshift for supernovae Ia progenitor, quasars, and gamma-ray bursts.

## 6. DISCUSSION

The advantage of considering quasars and gamma-ray bursts for measuring cosmological distances and testing cosmological models as compared to the type Ia supernovae is that they greatly extend the cosmological time scale from the cosmic time corresponding to $z \sim 1.5$ for supernovae to $z$ up to 10 for quasars and GRBs. The differences in the redshift coverage mean that while supernovae cover the epoch of $\Lambda$ dominance, quasars and GRBs include the epoch when $\Omega_M$ ruled the expansion of the Universe. However, currently, the statistical errors and scatter in the quasar and GRB data are too large compared to the supernovae data to consider them reliable standard candles for distance measurement. Therefore, one uses the cosmological parameters determined from SNeIa data to calibrate quasars and GRBs data and see how well the calibrated data fit the quasar and GRB observations statistically.

In their recent paper, Lusso et al. [87] wrote: 'We confirm that, while the Hubble diagram of quasars is well reproduced by a standard flat $\Lambda$CDM model (with $\Omega_M = 0.3$) up to $z \sim 1.5$, ...., a statistically significant deviation emerges at higher redshifts, in agreement with our previous works [e.g., 88, 89, 90] and other works on the same topic [e.g., 91].' They then tried to fit the quasar Hubble diagram with a flat $w_0w_a$CDM model, which is a commonly used extension of the standard $\Lambda$CDM model, where the parameter $w$ of the equation of state of the dark energy is considered to vary with redshift $z$ according to the parametrization $w(z) = w_0 + w_a \times (1-a)$, with $a = (1+z)^{-1}$ being the scale factor. It essentially introduces two extra parameters, $w_0$ and $w_a$ for fitting the Hubble diagram data without cross-checking for their correctness or validity.

In the CCC approach, we basically have two parameters ($q$ and $\beta$) representing the mass evolution of the emitters, other than the cosmological parameters ($H_0$, $\Omega_{m,0}$, and $\Omega_{\Lambda,0}$) to fit the quasar and GRB data. However, for quasars, we have tried to validate these parameters from the observed variation of the quasar black hole masses with redshift. We see that the parameters $q = -7.462^{-6.038}_{-8.885}$ and $\beta = 1.046^{1.3730}_{0.7196}$ determined from fitting their Hubble diagram, and $q = -7.33^{-6.149}_{-8.510}$ and $\beta = 0.8424^{1.149}_{0.5361}$ determined from fitting the quasar mass data, match within their 95% confidence bounds. For GRBs, we do not have any measured data on their mass evolution with redshift to compare the values we obtained by fitting the Hubble data. The strong degeneracy between $q$ and $\beta$ values by fitting the data, and a relatively flat minimum of $\chi^2$ values near $q = 1$, lead us to fix $q$ value to unity and then determine $\beta = 0.3315$ ($\pm 0.0865$). It means we are essentially predicting that the GRB mass evolves with redshift as $\ln(M_{grb}/M_{grb,0}) = ((1+z)^{1/3} - 1)$. Such mass variation should be considered for calibrating GRBs as standard candles. As discussed under *Results* of Sec. 5 in the context of GRBs, we believe that the CCC model provides a significantly better fit over the full range of the redshift than does the $\Lambda$CDM model. Thus, the CCC model may be better when using GRBs as standard candles.

As we have consistently good data fit using the CCC model, the function $f = \exp[(1+z)^{-1.8} - 1] = \exp(a^{1.8} - 1)$ can be considered to reasonably represent the coupling constants' variation (Equation 34). This variation then leads to $\dot{G}/G = 5.4a^{1.8}\dot{a}/a = 5.4a^{1.8}H$. At the current time (i.e., $z = 0$ or $a = 1$), $(\dot{G}/G)_0 = 5.4H_0$, and since $H_0 = 70.83(\pm 0.66)$ km s$^{-1}$ Mpc$^{-1}$ = 7.23 ($\pm 0.07$) $\times 10^{-11}$ $yr^{-1}$, we get $(\dot{G}/G)_0 = 3.90(\pm 0.04) \times 10^{-10}$ $yr^{-1}$. The confidence bound for this value is the same as for $H_0$, i.e., 95%. This constraint on $(\dot{G}/G)_0$ is shown in Table 2, along with some others determined by various methods. One immediately notices that the constraint determined by the CCC approach is among the most relaxed. It is primarily due to the concurrent variation of several coupling constants treated in the CCC approach. All other methods ignore the potential variation of constants other than the gravitational constant. In the cases where the evolution of stellar objects is involved on cosmological time scales [e.g., 2, 6, 7, 8, 11], there is an additional concern related to energy conservation. It is because energy is not conserved in cosmological evolution and general relativity [35, 92, 93, 94, 95, 96].

Since variations of $c$, $G$, $h$, and $k$ are interrelated as $G \sim c^3 \sim h^3 \sim k^{3/2}$, the constraint we have determined for $G$ also applies to $c$, $h$, and $k$ through $\dot{c}/c = \dot{h}/h = 1/2\,\dot{k}/k = 1/3\,\dot{G}/G$. One should, therefore, determine the constraint on the variation of the dimensionless function $f(z)$ rather than on the variation of any of the constants.

From Noether's theorem point of view, one may see temporal violation of *symmetry* associated with the coupling constants. However, a new constant $\alpha = 1.8$ now appears through Equation (53) which could be considered a fundamental constant of nature.

However, our method of constraining the coupling constants variation is indirect, i.e., derived rather than measured. A Kibble balance measures the gravitational mass of a test mass with extreme precision [97] by balancing the gravitational pull on the test mass against the electromagnetic lift force, resulting in equations leading to mass measurements involving the coupling constants. We are collaborating with NIST (National Institue of Standards and Technology, USA) to study the possibility of using the Kibble balance measurement of a test mass over several years to see if it could confirm or falsify our predictions of the coupling constants' variation.

TABLE 2 Constraints on $(\dot{G}/G)_0$ determined by various methods. CL means 'confidence limit', and NA means 'not available'.

| Look back time in years | Method | $(\dot{G}/G)_0 \times 10^{12}$ High limit | Low limit | CL | Reference |
|---|---|---|---|---|---|
| ~40 | Helioseismology | <0.2 | | 95% | Bonanno & Fröhlich [98] |
| ~45 | Lunar Laser Ranging | 0.147 | -0.005 | NA | Hofmann & Müller [12] |
| ~45 | Planetary Ephemeris | 0.078 | -0.07 | 95% | Pitjeva & Pitjev [99] |
| ~60 | White Dwarf Pulsations | 40 | -250 | 95% | Benvenuto et al [100] |
| ~3000 | Pulsar Timing | 9 | -27 | 68% | Kaspi et al [101] |
| ~100 Million | Pulsar Mass | 3.6 | -4.8 | 95% | Thorsett [8] |
| ~200 Million | Gravitational Waves | 20000 | -4000 | 90% | Vijaykumar et al. [16] |
| ~4 Billion | Young Sun Luminosity | | -4 | NA | Sahini & Shtanov [3] |
| ~5 Billion | Supernovae Type Ia | 10 | NA | NA | Gaztanaga et al [55] |
| ~5 Billion | White Dwarf Cooling | | <-50 | NA | Althaus et al. [6] |
| ~10 Billion | Stellar Astroseismology | <=5.6 | | 95% | Bellinger & C.-Dalsgaard [11] |
| ~10 Billion | Age of Globular Clusters | 7 | -35 | NA | Degl'Innocenti [7] |
| ~13 Billion | CCC – SNeIa, Quasars, GRBs | 394 | 386 | 95% | This paper |
| ~13 Billion | Cosmic Microwave Background | 1.05 | -1.75 | 95% | Wu & Chen [98] |
| ~14 Billion | Big Bang Nucleosynthesis | 4.5 | -3.6 | 95% | Alvey et al. [10] |

## 7. CONCLUSION

Astrophysicists have been concerned about the inability of the ΛCDM model to adequately calibrate the observation of quasars and gamma-ray bursts for their use as high redshift standard candles. We have shown that the Hubble diagrams of xA quasars and certain types of GRBs calibrated using supernovae type Ia data can test cosmological models up to $z\sim10$. Our covarying coupling constants approach fits the Hubble diagrams data admirably well over the full range of observations using the same cosmological parameters that fit the Hubble diagram of SNeIa from the Pantheon data. We, therefore, believe that the CCC model may be better than the ΛCDM model when using quasars and GRBs as standard candles. Since the variation of coupling constants is interrelated in the CCC model as $G \sim c^3 \sim h^3 \sim k^{3/2}$ through the function $f(a) = \exp(a^{1.8} - 1)$, we can express the general constraint in terms of the Hubble constant $H_0$ as $(\dot{G}/G)_0 = 3(\dot{c}/c)_0 = 3(\dot{h}/h)_0 = 1.5(\dot{k}/k)_0 = 5.4H_0 = 3.90\,(\pm 0.04) \times 10^{-10}$ yr$^{-1}$. The CCC model applied to SNeIa predicts their luminosity to increase with the redshift due to an increase in Chandrasekhar mass of white dwarfs. The model also predicts and verifies the increase of quasar mass and, concomitantly their luminosity with the redshift. Finally, the model predicts the mass (and luminosity) of gamma-ray bursts to increase with redshift as $\ln(M_{grb}/M_{grb,0}) = ((1+z)^{1/3} - 1)$.

## Acknowledgments

The author is grateful to Dr. Puxun Wu, Dr. Celia Escamilla-Rivera, Dr. Paola Marziani, Dr. Guido Risaliti, and Dr. Susanna Bisogni for providing the data used in this work. He is also thankful to Dr. Maria Dainotti for her correspondence and sharing data files related to her 3D Fundamental Plane for calibrating GRB data. He acknowledges an unconditional grant from Macronix Research Corporation in support of the research. Spcial thanks are due to the anonymous reviewers of the paper manuscript for their constructive critical comments for improving the quality and clarity of the paper.

## Data Availability

All the data used in this research is available from the cited references.

## Conflict of Interest

This work is performed without any conflict of interest.